\documentclass[12pt]{spieman}  
\usepackage{amsmath,amsfonts,amssymb}
\usepackage{graphicx}
\usepackage{setspace}
\usepackage{tocloft}
\usepackage{algorithm}
\usepackage{algpseudocode}
\usepackage{amsmath}
\usepackage{relsize}
\usepackage{multirow}

\title{Information Processing in Hybrid Photonic Electrical Reservoir Computing}

\author[a,*]{Prabhav Gaur}
\author[a]{Chengkuan Gao}
\author[a]{Karl Johnson}
\author[a]{Shimon Rubin}
\author[a]{Yeshaiahu Fainman}
\author[a]{Tzu-Chien Hsueh}
\affil[a]{University of California San Diego, Department of Electrical and Computer Engineering, 9500 Gilman Drive, La Jolla, CA 92092, USA}

\cftpagenumbersoff{figure}
\cftpagenumbersoff{table} 
\begin{document} 
\maketitle

\begin{abstract}
Physical Reservoir Computing (PRC) is a recently developed variant of Neuromorphic Computing, where a pertinent physical system effectively projects information encoded in the input signal into a higher-dimensional space. 
While various physical hardware has demonstrated promising results for Reservoir Computing (RC), systems allowing tunability of their dynamical regimes have not received much attention regarding how to optimize relevant system parameters. In this work we employ hybrid photonic-electronic (HPE) system offering both parallelism inherent to light propagation, and electronic memory and programmable feedback allowing to induce nonlinear dynamics and tunable encoding of the photonic signal to realize HPE-RC.  Specifically, we experimentally and theoretically analyze performance of integrated silicon photonic on-chip Mach-Zehnder interferometer and ring resonators with heaters acting as programmable phase modulators, controlled by detector and the feedback unit capable of realizing complex temporal dynamics of the photonic signal. Furthermore, we present an algorithm capable of predicting optimal parameters for RC by analyzing the corresponding Lyapunov exponent of the output signal and mutual information of reservoir nodes. By implementing the derived optimal parameters, we demonstrate that the corresponding resulting error of RC can be lowered by several orders of magnitude compared to a reservoir operating with randomly chosen set of parameters.
\end{abstract}


{\noindent \footnotesize\textbf{*}Prabhav Gaur,  \linkable{pgaur@ucsd.edu} }

\begin{spacing}{2}   

\section{Introduction}
\label{sect:intro}  
Reservoir Computing (RC) \cite{LUKOSEVICIUS2009127,Gauthier2021} is a variant paradigm within the field of Recurrent Neural Network (RNN), offering an effective approach for solving complex tasks, particularly in the domain of time-series prediction and signal processing. \cite{PhysRevLett.120.024102,VERSTRAETEN2007391}. A distinctive feature of RC is that in contrast to traditional RNN architectures where the entire network is trained end-to-end, RC employs a typically large reservoir composed of a random and fixed network of connected nodes which is not subject to training during the learning process, and a digitally trainable readout layer. The reservoir serves as a dynamic memory system that processes input signals by propagating them through its interconnected nodes and fits well into the PRC paradigm which usually has a fixed dynamics and reservoir structure. The key property of the reservoir is its echo-state property,\cite{GRIGORYEVA2018495} which means that its internal dynamics preserves a memory of the input signals over time. The input signals are fed into the reservoir, where they undergo nonlinear transformations as they propagate through the network’s recurrent connections. These transformed signals create complex temporal representations within the reservoir’s dynamics, capturing relevant features of the input data while preserving temporal dependencies. Following the reservoir’s processing of input signals, the output is extracted through the linear readout layer, which is a trainable component responsible for mapping the reservoir’s dynamic states to the desired output. RC has demonstrated impressive performance across various domains, including speech recognition \cite{1716215}, financial forecasting \cite{10.1371/journal.pone.0246737}, time-series prediction\cite{Chen2022}, and pattern recognition \cite{8545471}. 
RCs have been realized in numerous physical systems \cite{TANAKA2019100} including optical systems. While the latter offers unique advantages like high connectivity and wavelength multiplexing, and furthermore led to realization of several free-space and on-chip RC architectures, purely photonic RC (at least in the linear regime where nonlinear optical effects do not emerge) would only admit nonlinearity on the detector layer \cite{Tait2017, 8364605,VanderSandeBrunnerSoriano+2017+561+576,Wetzstein2020,Prucnal2017-kd,PhysRevX.7.011015,PhysRevX.10.041037,Gao2023}. 
One approach to introduce nonlinearity into photonic circuit without relying on nonlinear optical effects, is to employ Hybrid Photonic-Electronic (HPE) system that combines the advantages of both photonic and electronic components. 
In particular, Hybrid Photonic-

Electronic Reservoir Computing (HPE-RC)\cite{Nakajima2021,Kanno2022} is an innovative approach offering  speed, bandwidth, and parallelism inherent to photonic degrees of freedom, capable of generating complex temporal dynamics  and ease of integration with existing electronic systems\cite{McMahon2023} which in turn offer robust memory and nonlinear feedback mechanism. 
This hybridization enables HPE-RC to harness the strengths of both domains while mitigating their respective limitations leading to efficient, high-performance computing platforms with a wide range of applications \cite{Kumar:21}.

In particular, the mature  Photonic Integrated Circuits (PICs) technology \cite{Takano:18} enables to confine light in waveguides of submicron lateral dimensions, arranged in a complex network that enables to employ integrated modulators implementing nonlinear transformations of the input signal \cite{Vandoorne2014,Brunner:15,Sunada2019,Abdalla:23}, and hence highly conducive for efficient application of HPE-RC. The electronic hardware on the other hand is more efficient, in terms of cost and energy, providing feedback \cite{Duport:12} and furthermore employing field-programmable gate arrays (FPGAs) \cite{antonik2015fpga}, digital signal processors (DSPs)\cite{6782741}, or specialized hardware accelerators \cite{10.1145/2967446.2967447}, allows to match the speed of the input and the electrical readout layer. 
The latter is of particular importance as it allows to perform to process information useful for task-specific computation by mapping the reservoir's dynamic state onto the desired output. This mapping is usually done by linearly transforming the state value of the nodes via weights trained by various regression methods. \cite{FERNANDEZDELGADO201911} 

In fact, viable information processing and analysis employing concepts from dynamics theory is instrumental in realizing successful RC applications, where for instance determining Lyapunov exponents  \cite{WOLF1985285} from data via RC
was used to predict evolution of complex dynamical system \cite{PhysRevLett.120.024102}
The sign of Lyapunov exponent, which measures the rate of exponential divergence or convergence of nearby trajectories in the phase space, quantifies system's sensitivity to initial conditions where positive/negative sign leads to divergence/convergence of nearby trajectories and hence to chaotic/stable dynamics.
More recently, triggered by more fundamental questions investigating the relationship between Lyapunov exponents and the ability of reservoir to process information, some attention was given to quantify the performance of RC by considering the concept of mutual information \cite{doi:https://doi.org/10.1002/047174882X.ch2}, 
which quantifies to what extent the input signals is correlated with the reservoir's output responses.
Specifically, maximizing the mutual information allowed to enhance the performance of RC for both supervised \cite{Inubushi2017}
and unsuprevised learning regimes. \cite{TANAKA2020225} 
While recent works \cite{10416575,Zhang:23,Koster2023-cn,doi:10.1021/acsphotonics.4c00015,Guo2023,10412624,Ferreira_2022,9411652,10.1117/12.2544006,Harkhoe:20,Zhong:22}, demonstrated that photonic nonlinear systems can perform RC, to the best of our knowledge, there has been no study on demonstrating task-dependent optimization of HPE-RC system.
 
\begin{figure}
\begin{center}
\begin{tabular}{c}
\includegraphics[scale=1]{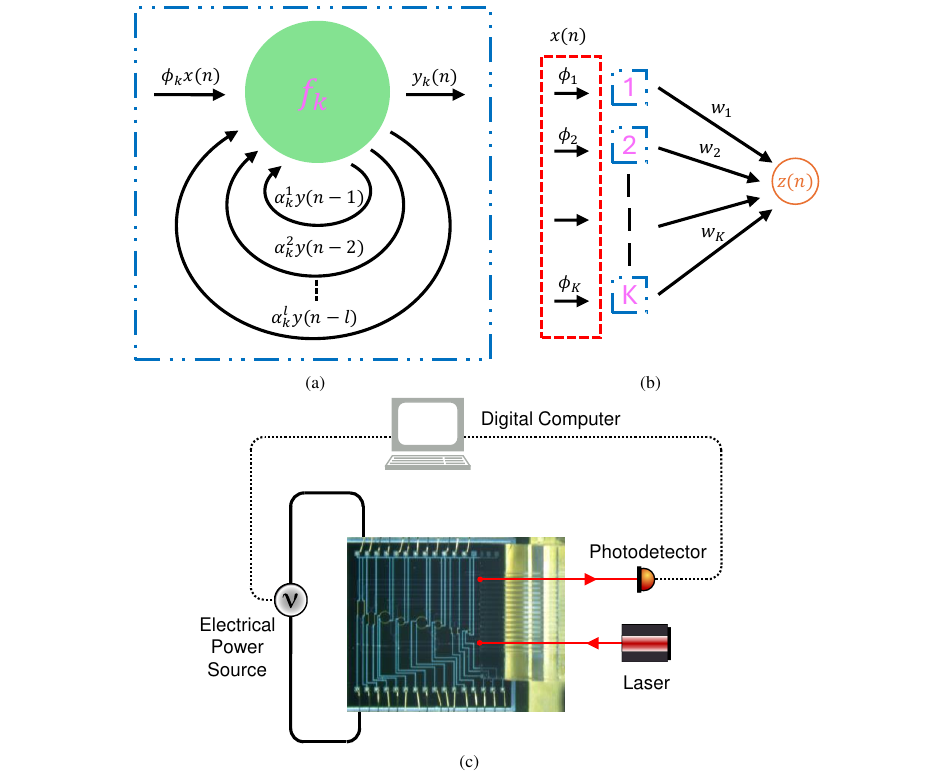}  
\end{tabular}
\end{center}
\caption 
{ \label{fig:reservoir}
Schematic description of the underlying principle of the implemented HPE-RC system in our work achieved with chip-scale Si photonic waveguides, MZI/ring resonator, CW laser, photodetector and integrated thermo-optical phase modulators controlled by electronic feedback system.
(a) Depiction of a single $k$\textsuperscript{th} node in the reservoir with a transmission function $f_k$ consisting of $\phi_k$ scale factor realized by modulator, and scaling of the output $y_k(n)$ by $\alpha^l_k$ depending on the delay length $l$, which is realized by electronic memory, using outputs at previous time steps as a feedback. The black arrrows are implemented electronically while the green circle is due to photonics design. (b) Formation of a reservoir comprising of $K$ parallel nodes with each node having its own characteristic parameters. The output from the nodes are trained using ridge regression and weights $w_k$ are calculated for each node to match the target function $z(n)$. (c) Setup to demonstrate on-chip HPE-RC. 
An optical fiber array is aligned with grating couplers on the chip and secured in place using index matching adhesive. 
A 1548 nm CW laser is connected to one fiber of the fiber array to supply optical power to the MZI through its respective input grating coupler. 
A photodetector is likewise connected to a different fiber in the array to detect light transmitted through the MZI, which is digitized and sent to the computer. The digital computer also controls the power source which drives current to the phase modulators on the optical devices on the chip forming the feedback loop of the system. The phase modulators are implemented using metal heaters which change the optical index of the arms of the MZI and the loop waveguide of the ring resonator. The learning of the weights in the final layer of HPE-RC is performed offline on the digital computer using ridge regression. }

\end{figure} 

In this work, we design and fabricate chip-scale silicon photonic HPE system and develop an algorithm capable to predict optimal HPE-RC performance by evaluating basic properties of signal dynamics and information correlations present in the system. Our hardware employs $220$ nm Silicon-on-Insulator (SOI) technology to fabricate MZI and Ring Resonator photonic devices fig~\ref{fig:reservoir}(c) each acting as a single node in HPE-RC, and are fed by a constant wave (CW) laser power, whereas the output is monitored by the photodetector and recorded by electronic memory allowing to realize feedback circuit (see more details in SM).
In particular, the photonic devices are supplemented with integrated metal heaters on the cladding layer which allow the feedback unit to modulate the phase of the waveguide mode via thermo-optic effect \cite{10.1063/1.4738989}, and in turn the output power.
The input and feedback are encoded in discrete time steps and the speed of the system is usually limited by the input/output hardware. Thus, each device acts like a node of the reservoir fig~\ref{fig:reservoir}(a), which has an input, a feedback and a nonlinear function operating on them to give an output. In this study, for simplicity, we assume that $K$ nodes present in the reservoir are not interlinked and each node processes information independently  fig~\ref{fig:reservoir}(b). Thus, for a unit laser power, the output from $k$\textsuperscript{th} node at discrete time step $n$ can be given by eq.~\ref{eq:1}

\begin{equation}
\label{eq:1}
y_k(n) = f_k\big(\phi_k x(n)+ \boldsymbol{ \alpha }_k \cdot \boldsymbol{ y}_k(n-1) + \theta_k 
\big),
\end{equation}
where $x(n)$ is the task-dependent input fed into the reservoir, $\phi_k$ is the input scaling factor for the $k$\textsuperscript{th} node, $\boldsymbol{y}_k(n-1) = \big[y_k(n-1),y_k(n-2),...,y_k(n-l)\big]^T$ is the feedback vector (which depends on the number of past delays used to provide the feedback $l$), $\boldsymbol{\alpha }_k = \big[\alpha_k^1, \alpha_k^2,...,\alpha_k^l\big]^T$ is the feedback scaling factor which forms a scalar product with $\boldsymbol{y}_k$, and $\theta_k$ is node's bias factor and can exist naturally due to the carrier wavelength or device structure, but can also be changed externally by adding a bias to the phase modulation.

The function $f_{k}$ is determined by the propagation of the photonic mode in the MZI device \cite{Chrostowski_Hochberg_2015} as a function of fixed geometric parameters and controllable phase modulators.
As phase modulation on an optical device is inherently nonlinear, a number of nodes with different $f_k$ can transfer input data to higher dimensions. 
Given a particular device and task, the intrinsic parameters $\phi$, $\boldsymbol{\alpha}$, and $\theta$ can be varied to form multiple nodes that process the input differently. To form a reservoir, multiple such nodes may be used in parallel, but owing to area/cost constraints only a limited number of nodes can be operated on a silicon chip. 
Thus, to form a reservoir with $K$ nodes, it becomes essential to choose the above parameters so that an optimal reservoir can be formed. 
For instance, if $K_p$ is the number of all parameters that can be implemented in a reservoir, then choosing  $K$ nodes from $K_p$ parameters can be realized in $\big(^{K_p}_K\big)$ combinations which is typically a large number, hence motivating the need for an approach enabling to predict optimal RC performance.

This work is structured as follows.
First we numerically demonstrate that optimal performance of HPE-RC can be achieved in a region of minimal Lyapunov exponent $\lambda(\alpha)$, considered as a function of feedback parameter $\alpha$. In particular we consider the performance of NARMA2 and NARMA10 series \cite{846741,PhysRevResearch.3.043135} considered as standard benchmark tests for RC \cite{Duport2016-wm} and study the relation of Lyapunov exponent and mutual information on reservor performance.
We calculate the Lyapunov exponent of these nodes as a function of the intrinsic parameters and study its relation to their performance as a reservoir. 
We then develop and implement the Lyapunov stable minimal redundancy maximal relevance (Ls-mRMR) algorithm which is an extension of the mRMR algorithm \cite{1453511} developed for feature selection in machine learning and experimentally validate the HPE-RC and Ls-mRMR on an integrated silicon photonic chip.

\section{Results}

\subsection{Lyapunov Exponent}

\begin{figure}
\begin{center}
\begin{tabular}{c}
\includegraphics[scale=1.0]{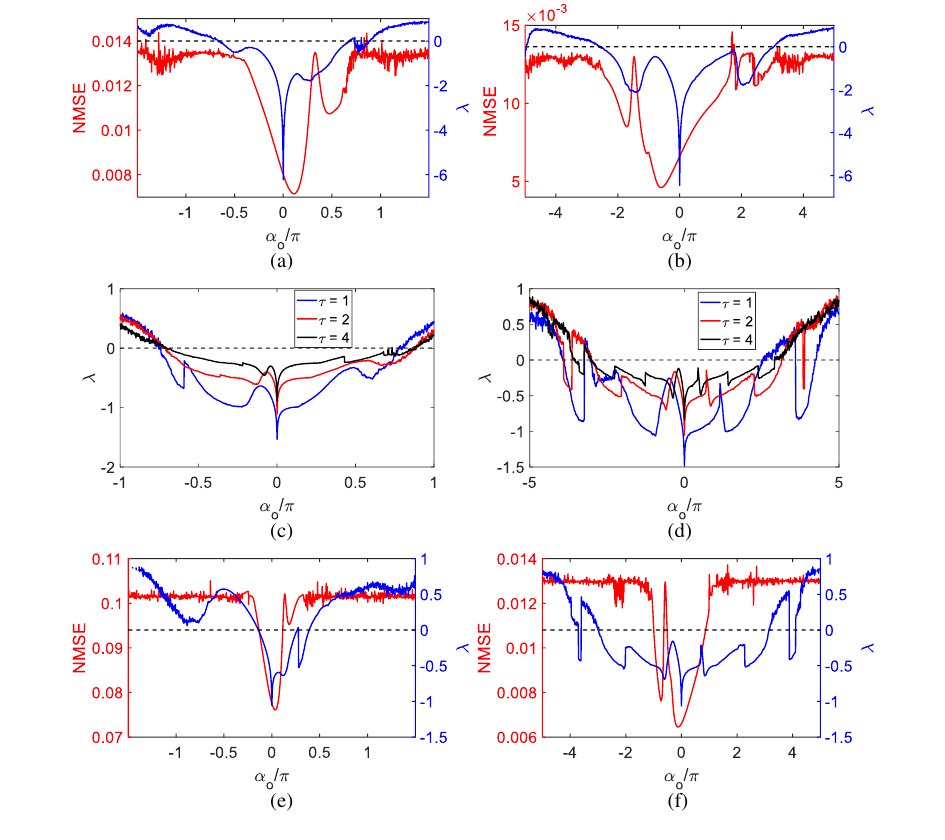}  
\end{tabular}
\end{center}
\caption 
{ \label{fig:LE}
but is still not too far from chaotic regime. 
Numerical results demonstrating correlation between NMSE of RC task and the value Lyapunov exponent as a function of the feedback parameter $\alpha_{0}$ and others parameters fixed; $\phi = -0.35\pi$ and $\theta = 0$. (a) Performance of MZI for NARMA2 with $l=1$ plotted against Lyapunov exponent. (b) Performance of ring resonator for NARMA2  with $l=1$ plotted against Lyapunov exponent. (c) Effect of $\tau$ on Lyapunov exponent of MZI for $l=9$. (d) Effect of $\tau$ on Lyapunov exponent of ring resonator for $l=9$. (e) Performance of MZI for NARMA10 with $l=9$ and $\tau = 2$ plotted against Lyapunov exponent. (f) Performance of ring resonator for NARMA10 with $l=9$ and $\tau = 2$ plotted against Lyapunov exponent.} 
\end{figure} 

We derive the equations for Lyapunov exponent from first principal (see Supplementary Material) and use it to develop the LE\_sim algorithm for calculating Lyapunov exponent of MZI and ring resonator PIC devices. In the following we drop the subscript in eq.~(\ref{eq:1}) representing $k$\textsuperscript{th} device, and treat $\eta(n) = \phi x(n)+\boldsymbol{ \alpha } \cdot \boldsymbol{ y}(n-1)+\theta$ as an argument for the device transfer function $f$, and $f'(\eta(n))$ stands for the corresponding derivative with respect to $\eta(n)$ at a particular time step $n$.
\begin{algorithm}
\caption{Lyapunov Exponent for simulation}\label{alg:cap}
\begin{algorithmic}
\Function{LE\_sim}{$f,\phi,\boldsymbol{\alpha},\theta$}
    \State $x \sim$ Normal distribution [0,1]
    \State $\boldsymbol{ y}(l-1) \gets \boldsymbol{0}_l$    ;   $S(0:l-2)\gets 0$   ;   $S(l-1) \gets 1$
    \For{$i=l:N$}
        \State $s(i) \gets f'(\eta(i))$
        \State $S(i) \gets s(i)\big[\boldsymbol{\alpha}\cdot S(i-1:-1:i-l)\big]$
    \EndFor
    \State $\lambda \gets \frac{1}{N}\ln(|S(N)|)$
    \State \textbf{return} $\lambda$
\EndFunction
\end{algorithmic}
\end{algorithm}
The algorithm is based on recursively calculating the derivative of the device function at each time step and averaging its logarithm over $N$ time step. 
In principle, most accurate calculation result is achieved when $N$ tends to infinity, however, for practical perspective the algorithm converges after few hundred time steps (we use $N$ upto 800 time steps). 
Also, large $N$ makes the value of $S(i)$ too large or too small beyond machine precision.
For simplicity, we assume negligible power loss in the optical device and the function\cite{Chrostowski_Hochberg_2015}  for MZI ($f_M$) and ring resonator ($f_R$) is given by eq.~(\ref{eq:2})
\begin{equation}
\label{eq:2}
f_M(\eta(i)) = \frac{1}{2}(1+\sin(\eta(i))) \text{   \quad  ;  \quad   } f_R(\eta(i)) = 
  \frac{4t^2\sin^2(\eta(i)/2)}{t^4 - 2\cos(\eta(i))t^2+1 }
\end{equation}
$t$ in the above equation is coupling coefficient of the ring with the bus waveguide and is assumed 0.45 throughout the simulations. To ensure fading memory in the electrical feedback we set $\boldsymbol{\alpha}$
\begin{equation}
\label{eq:3}
\boldsymbol{\alpha}=\alpha_o\exp{\Big(0:-\frac{1}{\tau}:-\frac{l-1}{\tau}   \Big)}
\end{equation}
$\alpha_o$ determines the feedback strength while $\tau$ determines the how quickly the memory fades. We study the relation between the performance of MZI as a single node of reservoir through the Normalized Mean Sqaure Error (NMSE)(see Supplemnatary Material) and its Lyapunov exponent. An example of such a comparison is shown in fig.~\ref{fig:LE} where we plot both the metrics as a function of $\alpha_o$  with fixed $\phi =-0.35\pi$ and $\theta=0$. The task performed is NARMA2 with $l=1$ and NARMA10 with $l=9$ and $\tau = 2$ for both MZI and ring resonator. We also compare the effect of $\tau$ on the Lyapunov exponent for both the devices in fig.~\ref{fig:LE}. As value of $\tau$ increases the memory of the system increases and the dynamic gets closer to chaotic regime ($\lambda>0$). It is evident from this result that the performance of the reservoir is worse when the reservoir operates in chaotic region. This is expected as the state much further behind in time greatly influences the current state of the reservoir and reservoir loses its echo state property. There are some sets of parameters in non-chaotic regimes that perform much better than others, but there is no further clear correlation between Lyapunov exponent and NMSE, although, the best performance is often around a maxima of the Lyapunov when its value is less than zero (see Supplementary Material).  Lyapunov exponent is a good metric to determine whether a set of parameters can be expected to perform better or worse for RC. But, after selecting a set of parameters based on Lyapunov exponent, it is not clear which subset will perform better for RC. Thus, to select $K$ parallel nodes from $K_p$ set of parameters we study the information present in the nodes of the RC.

\subsection{Mutual Information}

\begin{figure}
\begin{center}
\begin{tabular}{c}
\includegraphics[scale=1.0]{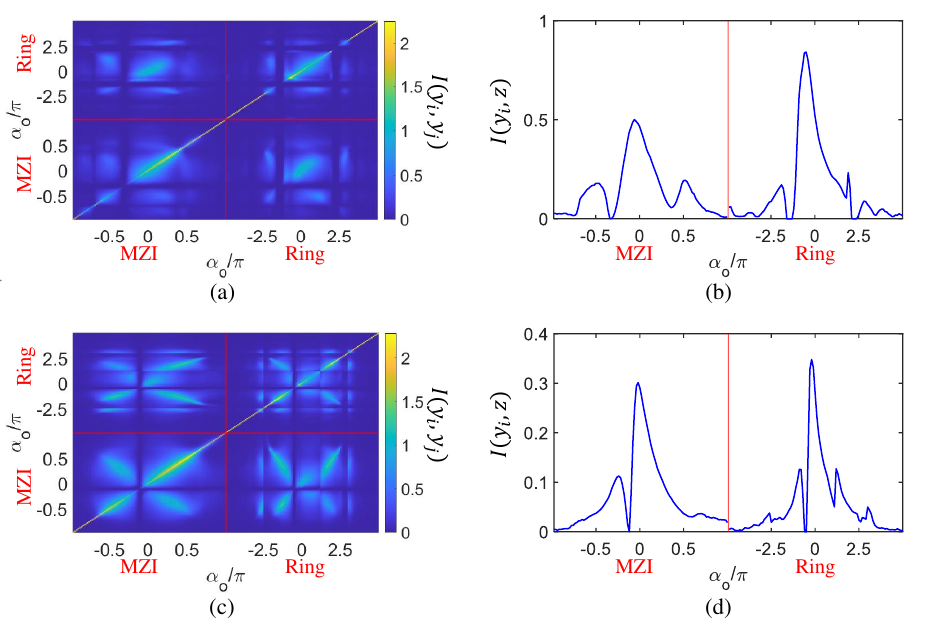}  
\end{tabular}
\end{center}
\caption 
{ \label{fig:LE}
Mutual information in the reservoir with $\tau = 2$, $\phi = -0.35\pi$ and $\theta = 0$ kept constant and $\alpha_o$ is varied. The mutual information depends on the parameters of the the nodes and the nodes need to be carefully chosen according to these values to form an optimal reservoir. (a) $I_R$ of MZI and ring resonator with $l=1$. (b) $I_D$ of MZI and ring resonator for NARMA2 with $l=1$. (c)  $I_R$ of MZI and ring resonator with $l=9$. (d) $I_D$ of MZI and ring resonator for NARMA2 with $l=1$. } 
\end{figure} 

Let the output from $i$\textsuperscript{th} and $j$\textsuperscript{th} node be $y_i(n)$ and $y_j(n)$. The output as a function of time can be used to determine their probability density function (p.d.f) $p(y_i)$ and $p(y_j)$ respectively, and also calculate their joint p.d.f $p(y_i,y_j)$. It should be noted that the output signals are analog in nature and need to be digitized before their p.d.f can be calculated. For simulation purposes, the precision of digital computers is too high to accurately determine the p.d.f with limited number of samples. For practical purposes, the precision of analog-to-digital converters (ADCs) can also be high too. Furthermore, it is computationally expensive to process p.d.f with high precision. Hence, in general, the output signal has to be binned accordingly to calculate an accurate p.d.f with limited samples. Using the p.d.f of the output, the mutual information between two parallel nodes, which is referred to as $I_R$, can be calculated by 
\begin{equation}
\label{eq:4}
I(y_i;y_j) = \mathlarger{\mathlarger{\sum_{y_j}}}\mathlarger{\mathlarger{\sum_{y_i}}} p(y_i,y_j)\log\bigg(\frac{p(y_i,y_j)}{p(y_i)p(y_j)}   \bigg)
\end{equation}
The higher the value of $I(y_i;y_j )$, the more information is present between the nodes. Let the target function that the reservoir has to emulate (NARMA models in this study) given by $z$. Then the mutual information between the $i$\textsuperscript{th} node and the desired output $z$, which is referred to as $I_D$ is given by 
\begin{equation}
\label{eq:5}
I(y_i;z) = \mathlarger{\mathlarger{\sum_{z}}}\mathlarger{\mathlarger{\sum_{y_i}}} p(y_i,z)\log\bigg(\frac{p(y_i,z)}{p(y_i)p(z)}   \bigg)
\end{equation}

The plot for mutual information for NARMA2 with $l=1$ and NARMA10 with $l=9$ is shown in fig.~\ref{fig:LE}. These values provide insight of how much information each node has with respect to other nodes and also the target function. We use the Lyapunov exponent and the mutual information to develop an algorithm to find an optimal reservoir in the next section.

\subsection{Lyapunov stable minimal redundancy maximal relevance algorithm}
The mutual information in eq.~\ref{eq:4} numerically provides a value of relevance ($I_R$) of one node for the RC task compared to other. Thus, if a node is being used for RC then a node with a lower value of $I_R$ is heuristically better to add to the reservoir as it contains a less redundant information. Similarly,  the mutual information in eq.~\ref{eq:5} numerically provides a value of dependence ($I_D$) of one node on target function. We would like to have nodes in a reservoir with higher value of $I_D$ as it contains more information of target output. As $\phi$, $\alpha_o$ and $\theta$ are analog parameters, they can form an infinite set in a given range. But due to physical constraints such DAC resolution, noise, etc., they can be implemented only with a finite resolution. And only a subset of these parameters can be implemented on-chip due to space and cost constraints. Let the set of parameters that can be implemented be $P^o$ with the elements $\{\phi^i,\alpha_o^i,\theta^i \} \in P^o$ and $|P^o |=K^o$ elements. The set of output from the nodes formed by this set of parameters can be given by mapping the $Y(.)$, and the set $Y(P^o)$ will also contain $K^o$ elements. We can eliminate a subset of parameters from this universal set $P^o$ as we know some parameters lead to chaotic nodes which is undesirable for RC. Thus, a subset $P\subseteq P^o$ with $K_p$ elements is chosen according to the value of Lyapunov exponent of the nodes. This helps in shrinking the search space and greatly reduce the time taken to implement the entire algorithm. The feature selection problem is to select a set $P_K$ with $K$ ($<K_p$) elements which is a subset of $P$ and its parameters create nodes that form an optimal reservoir for a given task. Thus, we have to maximize the following cost function 

\begin{equation}
\label{eq:6}
C(P_K,z) = \frac{1}{K}\mathlarger{\mathlarger{\sum}}_{y_i\in Y(P_K)} I(y_i;z)  - \frac{1}{K^2}\mathlarger{\mathlarger{\sum}}_{y_i,y_j\in Y(P_K)} I(y_i;y_j) 
\end{equation}

A brute force search without using Lyapunov exponent and mutual information to optimize the performance of reservoir would result in checking $\big(\begin{smallmatrix} K^o \\ K \end{smallmatrix}\big)$ combination, which result a computational complexity of $O(K^{oK} )$ and is practically impossible on a regular digital computing device. After reducing the number of parameters by using Lyapunov exponent, the computational complexity to directly optimize the performance of reservoir or do a global search to optimize eq.~\ref{eq:6} still remains $O(K_P^K )$. We optimize the reservoir using incremental search method, the Lyapunov stable minimal redundancy maximal relevance (Ls-mRMR) algorithm, which is an extension of feature selection criterion used in machine learning \cite{1453511}

\begin{algorithm}
\caption{Lyapunov stable minimal redundancy maximal relevance}\label{alg:Ls_mRMR}
\begin{algorithmic}
\Function{Ls\_mRMR}{$f,P^o,K,z$}
    \For{$\{\phi^i,\boldsymbol{\alpha}^i,\theta^i\}\in P^o$}
        \If{$\lambda_{min} < \text{LE\_Sim}(f,\phi^i,\boldsymbol{\alpha}^i,\theta^i)<0$}
            \State $P\ni\{\phi^i,\boldsymbol{\alpha}^i,\theta^i\}$
        \EndIf
    \EndFor
    \State \textbf{select} $P_1 \ni \{ \phi^1,\boldsymbol{\alpha}^1,\theta^1\}$ \textbf{s.t.} $P_1\subset P$
    \State \textbf{and} $  y_1 = Y(P_1) = \arg \max\limits_{y_i\in Y(P)} I(y_i;z) $
    \State $P\gets P - P_1$
    \For{$k=2:K$}
        \State \textbf{select} $D_k \ni \{ \phi^k,\boldsymbol{\alpha}^k,\theta^k\}$ \textbf{s.t.} $D_k\subset P$
        \State \textbf{and} $  y_k = Y(D_k) = \arg \max\limits_{y_i\in Y(P)} \big[ I(y_i;z) - \frac{1}{k-1} \sum\limits_{y_j\in Y(P_{k-1})} I(y_i;y_j)\big]$
        \State $P_k \gets P_{k-1}+D_k $
        \State $P \gets P-D_k $
    \EndFor    
    
    \State \textbf{return} $P_K$
\EndFunction
\end{algorithmic}
\end{algorithm}

The above algorithm is based on incremental  search where in each iteration a new node is selected to be added to a set of already selected nodes. The computational complexity of this algorithm is $O(K K_p )$ which can be easily implemented digitally. We simulate the working of the algorithm for both MZI and ring resonator. We linearly divide the parameters \{$\phi,\alpha_o,\theta$\} in \{6,5,6\} segments in ranges $\{  [-\pi/2 \text{ } \pi/2], [\pi \text{ } \pi], [-\pi/2 \text{ } \pi/2]  \}$ and $\tau=2$ is kept constant. Thus, we get a set $P^o$ with $K^o=180$. We would like to choose $K=10$ parameters to build a reservoir. Additive white gaussian noise (AWGN) is added in all simulation such that source-to-noise ratio (SNR) is 1000. 1600 samples are used for training and 600 samples are used for testing. The result for NARMA2 and $l=1$ is shown in Table~\ref{tab:algo}. The performance of reservoir is measured by calculating the normalized mean square error (NMSE) of the testing phase. The blind error is calculated by randomly selecting 10 different set of parameters and averaging their NMSE for the task.  Similarly, for NARMA10 and $l=9$ performance is given by Table~\ref{tab:algo}.  As the nodes are more stable for higher delay length, the ranges for parameters \{$\phi,\alpha_o,\theta$\} used is $\{  [-\pi \text{ } \pi], [\pi \text{ } \pi], [-\pi \text{ } \pi]  \}$. The tables show the improvement that Ls-mRMR algorithm introduces for selecting the nodes of the reservoir compared to blind selection of parameters. Also, having the option to use both MZI and ring resonators improves the performance of the reservoir further, or alternatively, the more variety of device function that can be implemented the better will be the performance.

\begin{table}[ht]
\caption{Performance of reservoir for NARMA2 and NARMA10 task when $K=10$ nodes are selected with only MZI, only ring resonator and both MZI and ring resonator. The NMSE for using algorithm and selecting nodes blindly is compared.} 
\label{tab:algo}
\begin{center}       
\begin{tabular}{|c|c|c|c|c|} 
\hline
\rule[-1ex]{0pt}{3.5ex}  & & MZI & Ring & MZI \& Ring  \\
\hline\hline

\rule[-1ex]{0pt}{3.5ex}  \multirow{ 2}{*}{NARMA2} & Algorithm Error & $7.24$E-4 & $8.26$E-4 & $5.90$E-4  \\

\rule[-1ex]{0pt}{3.5ex}  & Blind Error & $1.25$E-2 & $1.25$E-2 & $1.22$E-2  \\
\hline
\rule[-1ex]{0pt}{3.5ex}  \multirow{ 2}{*}{NARMA10} & Algorithm Error & $5.14$E-2 & $4.52$E-2 & $4.44$E-2 \\

\rule[-1ex]{0pt}{3.5ex}  & Blind Error & $8.60$E-2 & $8.60$E-2 & $8.53$E-2 \\

\hline
\end{tabular}
\end{center}
\end{table} 

\subsection{HPE-RC on silicon photonics chip}

\begin{figure}
\begin{center}
\begin{tabular}{c}
\includegraphics[scale=1.0]{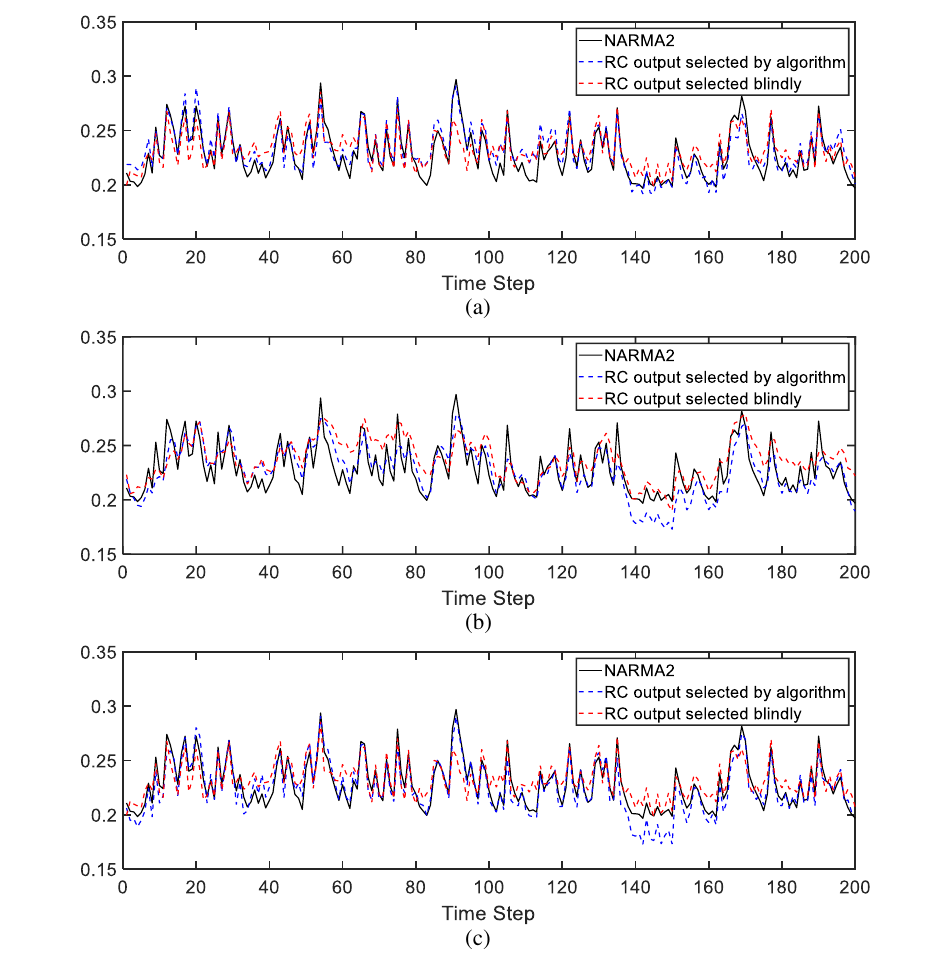}  
\end{tabular}
\end{center}
\caption 
{ \label{fig:chip}
Comparison of output of experimental RC against NARMA2 for feedback length $l=1$. (a) MZI as nodes of HPE-RC. (b)  Ring resonator as nodes of HPE-RC. (c) Both MZI and ring resonator as nodes of HPE-RC  } 
\end{figure} 

To demonstrate the HPE-RC experimentally, a sample was fabricated at a commercial silicon photonics foundry (Applied Nanotools Inc). An optical fiber array is aligned with grating couplers on the chip and secured in place using index matching adhesive. A 1548 nm CW laser is connected to one fiber of the fiber array to supply optical power to the MZI through its respective input grating coupler. A photodetector is likewise connected to a different fiber in the array to detect light transmitted through the MZI, which is digitized and sent to the computer. The digital computer also controls the power source which drives current to the phase modulators on the optical devices on the chip forming the feedback loop of the system. The phase modulators are implemented using metal heaters which change the optical index of the arms of the MZI and the loop waveguide of the ring resonator. The learning of the weights in the final layer of HPE-RC is performed offline on the digital computer using ridge regression.
The chip was wire bonded (Advanced International Technology, San Diego) and packaged with a 16-channel fiber array (Chip-scale Photonics Testing facility, UC San Diego). A 1548 nm continuous wave (CW) laser (Gooch and Housego AA1408) serves as the light source, and a source-meter power supply (Keithley 2400) is used to drive current through the metal heaters. A diagram of the setup is shown in fig.~\ref{fig:reservoir}c. It is often difficult to determine the exact transfer function of the device fabricated on chip. Thus, a different algorithm is needed to determine the Lyapunov exponent of each device. For a given input and a particular device the output $y(n)$ can be experimentally measured at each time step. Let $\gamma_o$ be a small perturbation given to the system. The Lyapunov exponent can then be calculated by a multidimensional Benettin Algorithm \cite{sprott2003chaos}.

\begin{algorithm}
\caption{Multidimensional Benettin Algorithm}\label{alg:MBA}
\begin{algorithmic}
\Function{LE\_Benettin}{$\boldsymbol{y}$}
    \State $x \sim$ Normal distribution [0,1]
    \State $\boldsymbol{ y}^1(l-1) \gets \boldsymbol{0}_t$    
    \State $\boldsymbol{ y}^2(l-1) \gets \boldsymbol{0}_t$      \State $y^2(l-1) \gets \gamma_o$
    \For{$i=l:N$}
        \State \textbf{Determine experimentally} $\boldsymbol{ y}^1(i), \boldsymbol{ y}^2(i)$
        \State $\gamma_i \gets \parallel \boldsymbol{ y}^2(i) - \boldsymbol{ y}^1(i) \parallel$
        \State $\boldsymbol{ y}^2(i) \gets \boldsymbol{ y}^1(i) + (\gamma_o/\gamma_i)(\boldsymbol{ y}^2(i)-\boldsymbol{ y}^1(i))$
    \EndFor
    \State $\lambda \gets \big<\ln\big(\frac{\gamma_k}{\gamma_o}\big)\big>_k$
    \State \textbf{return} $\lambda$
\EndFunction
\end{algorithmic}
\end{algorithm}
Now, the data from various devices can be collected by using different parameters similar to previous section and Ls-mRMR algorithm can be used to determine an optimal reservoir. The performance of reservoir for NARMA2 and $l=1$ is shown in table 3. The output of the reservoir during testing for MZI and ring resonator, both separate and together, is shown in fig.~\ref{fig:chip}. Thus, using an algorithm to choose the nodes for reservoir performs significantly better than choosing the nodes blindly. 

\begin{table}[ht]
\caption{Performance of on chip HPE-RC for NARMA2 task when $K=10$ nodes are selected with only MZI, only ring resonator and both MZI and ring resonator. The NMSE for using algorithm and selecting nodes blindly is compared.} 
\label{tab:algo}
\begin{center}       
\begin{tabular}{|c|c|c|c|} 
\hline
\rule[-1ex]{0pt}{3.5ex}   & MZI & Ring & MZI \& Ring  \\
\hline\hline

\rule[-1ex]{0pt}{3.5ex}   Algorithm Error & $1.65$E-3 & $3.61$E-3 & $1.58$E-3  \\

\rule[-1ex]{0pt}{3.5ex}   Blind Error & $4.17$E-3 & $9.42$E-3 & $4.17$E-3  \\
\hline

\hline
\end{tabular}
\end{center}
\end{table}

\section{Discussion}

Physical reservoir computing can be used to perform a number of machine learning tasks but there is no systematic approach to build a reservoir from hardware. We propose a novel method to assemble the nodes of an HPE-RC by studying the information present in the nodes. 
In this work we show that Lyapunov exponent can be correlated with the performance of the reservoir. The nodes that perform better lie in a certain region which is neither too far nor too close to the chaotic dynamics. 
The information present in the nodes of the reservoir is used to develop an algorithm that improves the performance of reservoir many-fold. As the parameters that are manipulated for optimization can be controlled electronically, it allows the reservoir to be reprogrammable for different tasks. We use digital computers and metal heaters to implement the parameters and provide the feedback. Although, being convenient, they increase the latency of the feedback, hence reducing the processing speed of reservoir computing. Electro-optic phase modulators with FPGA are better options to build the reservoir as they can make the reservoir operate as fast as I/O hardware and optimize the speed. Also, analog circuits on a different or same chip can reduce the area of the device. The ability to fabricate more device functions would mean that the algorithm can choose from a greater variety of features and improve the performance of the reservoir even further. MZI, ring resonator and other optical devices can be interconnected to create new device functions that can be used for reservoir computing. Optics also allow processing of signal at different wavelengths together. Wavelength division multiplexing can be used to implement several nodes at once increasing the performance of reservoir without increasing the physical area of the reservoir. Further study in these areas is needed to improve the performance of the reservoir even more.


\bibliography{report}   
\bibliographystyle{spiejour}   


\vspace{2ex}\noindent\textbf{Prabhav Gaur} received his B. Tech degree in Electrical Engineering from the Indian Institute of Technology Kanpur, India in 2019. He received his M.S. Degree in Electrical and Computer Engineering (Photonics Major) from the University of California San Diego in 2021. He is currently a Ph.D. candidate in the Department of Electrical and Computer Engineering at the University of California San Diego. His research interests include optical signal processing and optical metrology.

\vspace{2ex}\noindent\textbf{Karl Johnson} received his B.S. in Electrical Engineering from UC San Diego in 2023 with specializations in photonics and semiconductor devices. He is currently pursuing a Ph.D. in Electrical Engineering in the Photonics program at UC San Diego. His current research interests include chip-scale optical sensors and spectrometers, photonic MEMS, and packaging for integrated photonics.

\vspace{2ex}\noindent\textbf{Yeshaiahu Fainman} is ASML/Cymer Chair Professor and Distinguished Professor in ECE at UCSD. He received M. Sc and Ph. D degrees from Technion-Israel Institute of Technology in 1979 and 1983, respectively. He and his group made significant contributions to near field optical phenomena, nanophotonics, meta-materials, and plasmonics. He contributed over 340 manuscripts and over 560 conference papers. He is a Fellow of the OPTICA (former OSA), Fellow of the IEEE, Fellow of the SPIE, and a recipient of the Lady Davis Fellowship, Brown Award, Gabor Award, Emmett N. Leith Medal, Joseph Fraunhofer Award/Robert M. Burley Prize and OPTICA Holonyak Award.

\vspace{2ex}\noindent\textbf{Tzu-Chien Hsueh} received the Ph.D. degree in electrical and computer engineering from the University of California, Los Angeles, CA, in 2010. From 2001 to 2006, he was a Mixed-Signal Circuit Design Engineer in Hsinchu, Taiwan. From 2010 to 2018, he was a Research Scientist in Intel Lab Signaling Research and an Analog Engineer in Intel I/O Circuit Technology, Hillsboro, Oregon. Since 2018, he has been an Assistant Professor in electrical and computer engineering at the University of California, San Diego (UCSD). His research interests include wireline electrical/optical transceivers, clock-and-data recovery, data-conversion circuits, on-chip performance measurements/analyzers, and digital/mixed signal processing techniques. Dr. Hsueh was a recipient of multiple Intel Division and Academy Awards from 2012 to 2018, the 2015 IEEE Journal of Solid-State Circuits (JSSC) Best Paper Award, the 2020 NSF CAREER Award, and the 2022 UCSD Best Teacher Award. He served on the Patent Committee for Intel Intellectual Property (Intel IP) and the Technical Committee for Intel Design \& Test Technology Conference (DTTC) from 2016 to 2018. Since 2017, he has seen an IEEE senior member and served on the Technical Program Committee for IEEE Custom Integrated Circuits Conference (CICC) and the Guest Associate Editor for IEEE Solid-State Circuits Letters (SSC-L)

\vspace{1ex}
\noindent Biographies and photographs of the other authors are not available.

\listoffigures
\listoftables

\end{spacing}
\end{document}